\documentstyle[a4,12pt,axodraw]{article}

\newcommand{\sm}{standard model}
\newcommand{\mssm}{minimal supersymmetric \sm}
\newcommand{\lsp}{lightest supersymmetric particle}
\newcommand{\xs}{cross section}
\newcommand{\br}{branching ratio}
\newcommand{\EM}{electromagnetic}

\newcommand{\vev}{vacuum expectation value}

\newcommand{\rg}{renormalization group}
\newcommand{\nwa}{narrow width approximation}

\newcommand{\susic}{supersymmetric}
\newcommand{\susy}{supersymmetry}

\newcommand{\sel}{selectron}
\newcommand{\trm}{transverse momentum}
\newcommand{\cms}{center of mass system}
\newcommand{\cm}{center of mass}
\newcommand{\Sel}{\mbox{$\tilde e$}}
\newcommand{\snu}{sneutrino}
\newcommand{\Snu}{\mbox{$\tilde\nu$}}
\newcommand{\co}{chargino}
\newcommand{\Co}{\mbox{$\tilde\chi^\pm_1$}}
\newcommand{\no}{neutralino}
\newcommand{\No}{\mbox{$\tilde\chi^0_1$}}
\newcommand{\pT}{\mbox{$p_{\perp}$}}
\newcommand{\mpT}{\mbox{$p_{\perp}\!\!\!\!\!\!/\,\,\,$}}
\newcommand{\ep}{\mbox{$e^+e^-$}}
\newcommand{\GeV}{\mbox{ GeV}}
\newcommand{\GEV}{{\rm GeV}}

\newcommand{\hra}{\hspace{1mm}\hookrightarrow}

\def\lr3{$SU(3)_L\otimes SU(3)_R$}

\def\es{eigenstate}
\def\z0{$Z^0$}
\def\Z0{$Z^0$}

\def\ep{$e^+e^-$}

\def\cm{centre of mass}

\def\gsim{\buildrel{\lower.7ex\hbox{$>$}}\over{\lower.7ex\hbox{$\sim$}}}
\def\lsim{\buildrel{\lower.7ex\hbox{$<$}}\over{\lower.7ex\hbox{$\sim$}}}

\def\FB{{\rm fb}}
\def\PB{{\rm pb}}

\def\setepsfscale#1{\def\epsfsize##1##2{#1##1}}

\begin{document}

%  #] preamble:
%  #[ titlepage:
\thispagestyle{empty}

\setcounter{page}{0}

\begin{flushright}
CERN-TH.6742/92\\
MPI-Ph/92-73\\
LMU-92/07\\
October 1992
\end{flushright}
\vspace*{\fill}
\begin{center}
{\Large\bf Supersymmetric Signals in $\gamma\gamma$ Collisions}\\
\vspace{2em}
\large
\begin{tabular}[t]{c}
Frank Cuypers$^{a,1}$\\
Geert Jan van Oldenborgh$^{a,2,*}$\\
Reinhold R\"uckl$^{a,b,c,3}$\\
\\
{$^a$ \it Sektion Physik der Universit\"at M\"unchen,
D--8000 M\"unchen 2, FRG}\\
{$^b$ \it Max-Planck-Institut f\"ur Physik,
Werner-Heisenberg-Institut,}\\
{\it D--8000 M\"unchen 40, FRG}\\
{$^c$ \it CERN,
CH-1211 Gen\`eve 23, Switzerland}\\
\end{tabular}
\end{center}
\vspace*{\fill}

\begin{abstract}
We study the occurrence of final states with only an electron-positron pair
and missing \trm\ as a signal of \susy\ in photon-photon collisions.
Suitable high energy photon beams may be provided at linear colliders
by back-scattering laser beams on electron beams.
The final states considered
represent a typical signature for the production and decay
of \sel\ and \co\ pairs
within the \mssm.
We show that,
away from the kinematical threshold,
\sel s produce this signal
far more abundantly than \co s.
The standard model background is dominated by W-pair production.
We propose a series of kinematical cuts
which reduce this background to an acceptable level.
With a 1 TeV collider operated in the $\gamma\gamma$-mode,
we find that interesting and complementary tests
of \susic\ models
can be performed
for \sel\ masses up to 350 GeV.
\end{abstract}

\vspace*{\fill}

\begin{flushleft}
\noindent$^1$ {\small Email: {\tt frank@hep.physik.uni-muenchen.de}}\\
\noindent$^2$ {\small Email: {\tt gj@csun.psi.ch}}\\
\noindent$^3$ {\small Email: {\tt rer@dmumpiwh.bitnet}}\\
\noindent$^*$ Now at the Paul Scherrer Institut, CH-5232 Villigen PSI,
 Switzerland.\\
\end{flushleft}

\newpage

%  #] titlepage:
%  #[ introduction:

\section{Introduction}

Photon-photon collisions are a very attractive tool to search for
physics beyond  the \sm\
because the production rates for new hypothetical particles are
essentially known once their \EM\ charges are specified.
On the other hand,
new states are generally expected to be heavy
and can thus only be produced in high-energy collisions.
As a matter of fact,
it appears feasible to obtain suitable
energetic photon beams at linear colliders
by back-scattering a laser ray on an electron beam.

In this paper,
we show how an \ep\ linear collider operated in the $\gamma\gamma$ mode
can be used to probe and investigate \susy.
We focus on experimentally clean
and theoretically interesting processes:
the production and subsequent decay of pairs of
\sel s (\Sel) and \co s (\Co).
For our study
we assume the \mssm\ (MSSM).
In this case
there are two \co\ and four \no\ mass \es s
of which the lightest ones are denoted by
$\tilde\chi_1^\pm$ and $\tilde\chi_1^0$.
If R-parity is conserved,
as in the conventional MSSM,
the \lsp\ (LSP) is stable and escapes detection.
A likely candidate for the LSP is the lightest \no\ state \No.
Moreover,
the \sel\ and lightest \co\
are expected to decay into the channels
$\tilde e^\pm\to e^\pm\No$ and $\Co\to e^\pm\nu_e\No$,
which give rise to final states with only an electron-positron pair
and missing \trm.

Although the production \xs s are unambiguously predicted
as functions of the sparticle masses,
the \br s for the decay of
the \sel s and \co s
into the required final states
depend on further \susy\ parameters.
This dependence has to be carefully taken into account
when estimating the sensitivity limits for photon-photon collisions.
Conversely,
these parameters can be constrained if a positive signal is observed.

The $e^+e^-+\mpT$ signal should be compared to the background
resulting from several \sm\ processes,
dominantly from W-pair production
followed by the decay $W\to e\nu_e$.
We indicate the cuts necessary to suppress the background sufficiently
and to observe a potential signal.
Moreover,
we outline the range of \susy\ parameters accessible
in such an experiment.

Searches in $\gamma\gamma$ collisions are complementary to searches
in $e^+e^-$ \cite{LC1},
$e^-e^-$ \cite{LC11,LC18}
and $e^-\gamma$ \cite{LC10,LC16,LC19}
collisions
conducted at the same linear collider.
They provide useful cross-checks of signals which might be observed
in the latter reactions,
help to determine certain \susy\ parameters
and corroborate possible bounds.
Still,
the most important test of the \mssm\
remains the search for a light Higgs boson,
which is best performed in the $e^+e^-$ mode.

In the next section
we describe the relevant characteristics of the photon beams.
Then, in sections 3 and 4,
we compute the \xs s for the $e^+e^-+\mpT$ signal
from pair-production and decay
of \sel s and \co s.
Section 5 is devoted to the computation of the most dangerous \sm\ backgrounds.
In section 6
we investigate how to suppress the backgrounds by kinematical cuts.
Finally,
we compare the \susic\ signals with the irreducible background
and discuss the range of \susy\ parameters
which can be probed
at a 1 TeV collider.

%  #] introduction:
%  #[ photon beam:

\section{High-Energy Photon Beams}

High-energy photon beams
can be produced at linear colliders
by back-scattering a high-intensity
laser ray on a high-energy electron beam \cite{LC12}.
In principle,
if the laser intensity is high enough,
every electron in a bunch can interact
and yield a Compton photon.
With very powerful lasers
an electron can scatter more than once,
so the number of scattered photons
can even exceed the original number of electrons in the bunch.
Neglecting multiple scattering\footnote{This mainly enhances the low-energy
tail of the photon energy distribution (\ref{spectrum}),
which is irrelevant for the production of heavy (s)particles.}
and higher order effects
the resulting Compton photon energy spectrum is given by
\begin{equation}
P(y)=
{1\over N}
\left(
1-y+{1\over1-y}-{4y\over x(1-y)}+{4y^2\over x^2(1-y)^2}
\right)
\label{spectrum}
\end{equation}
where the factor $N$ normalizes the distribution to unity,
$\int_0^{y_{max}}dyP(y)=1$,
and
$y={E_{\gamma}/E_e}$
is the energy fraction of the electrons transferred to the photons.
It is bounded by
\begin{equation}
0\leq y\leq{x\over x+1}
\label{ylim}
\end{equation}
where
\begin{equation}
x={4E_eE_{laser}\over m_e^2}\leq2(1+\sqrt{2})\approx4.83~.
\label{xpression}
\end{equation}
The electron and laser beams are taken to be aligned
and their respective energies are
$E_e$ and $E_{laser}$.
For values of $x$ exceeding the upper bound imposed by (\ref{xpression}),
the laser and Compton photons can pair-produce \ep\ pairs
and the conversion efficiency drops dramatically.
In what follows,
we take $x=2(1+\sqrt{2})$.
Since both electron beams are to be converted,
the maximum attainable energy in the photon-photon \cms\ is thus
$\sqrt{s_{\gamma\gamma}}\leq2(\sqrt{2}-1)\sqrt{s_{ee}}\approx.83\sqrt{s_{ee}}$,
where $\sqrt{s_{ee}}=2E_e$.
Moreover, we assume the $\gamma\gamma$ luminosity
to be the same as the projected \ep\ luminosity
of the linear collider,
that is
${\cal L}_{\gamma\gamma}=10^{33}\div10^{34}~{\rm cm}^{-2}{\rm s}^{-1}$.

%  #] photon beam:
%  #[ selectrons:

\section{Production and Decay of Selectrons}

\begin{figure}[htb]
\begin{center}
\begin{picture}(300,150)(0,-30)
\Photon(00,00)(75,10){5}{5}
\Photon(75,90)(0,100){5}{5}
\Vertex(75,90){.5}
\DashLine(75,90)(75,10){5}
\Vertex(75,10){.5}
\DashLine(75,10)(150,0){5}
\DashLine(75,90)(150,100){5}
\Photon(200,00)(250,50){5}{5}
\Photon(250,50)(200,100){5}{5}
\Vertex(250,50){.5}
\DashLine(250,50)(300,0){5}
\DashLine(250,50)(300,100){5}
\put( 00,20){\makebox(0,0){$\gamma$}}
\put( 00,80){\makebox(0,0){$\gamma$}}
\put( 95,50){\makebox(0,0){$\tilde{e}$}}
\put(150,80){\makebox(0,0){$\tilde{e}^+$}}
\put(150,20){\makebox(0,0){$\tilde{e}^-$}}
\put( 75,-20){\makebox(0,0){+ crossed}}
\put(200,20){\makebox(0,0){$\gamma$}}
\put(200,80){\makebox(0,0){$\gamma$}}
\put(300,80){\makebox(0,0){$\tilde{e}^+$}}
\put(300,20){\makebox(0,0){$\tilde{e}^-$}}
\end{picture}
\\
\begin{picture}(300,150)(-80,-30)
\Photon(00,00)(75,10){5}{5}
\Photon(75,90)(0,100){5}{5}
\Vertex(75,90){.5}
\ArrowLine(75,90)(75,10)
\Vertex(75,10){.5}
\ArrowLine(75,10)(150,0)
\ArrowLine(150,100)(75,90)
\put( 00,20){\makebox(0,0){$\gamma$}}
\put( 00,80){\makebox(0,0){$\gamma$}}
\put( 95,50){\makebox(0,0){$\tilde{\chi}_1$}}
\put(150,80){\makebox(0,0){$\tilde{\chi}_1^+$}}
\put(150,20){\makebox(0,0){$\tilde{\chi}_1^-$}}
\put( 75,-20){\makebox(0,0){+ crossed}}
\end{picture}
\\
\end{center}
\caption[dummy]{Lowest order Feynman diagrams contributing to
                        \sel\ (top)
                        and \co\ (bottom) production.}
\label{f1}
\end{figure}
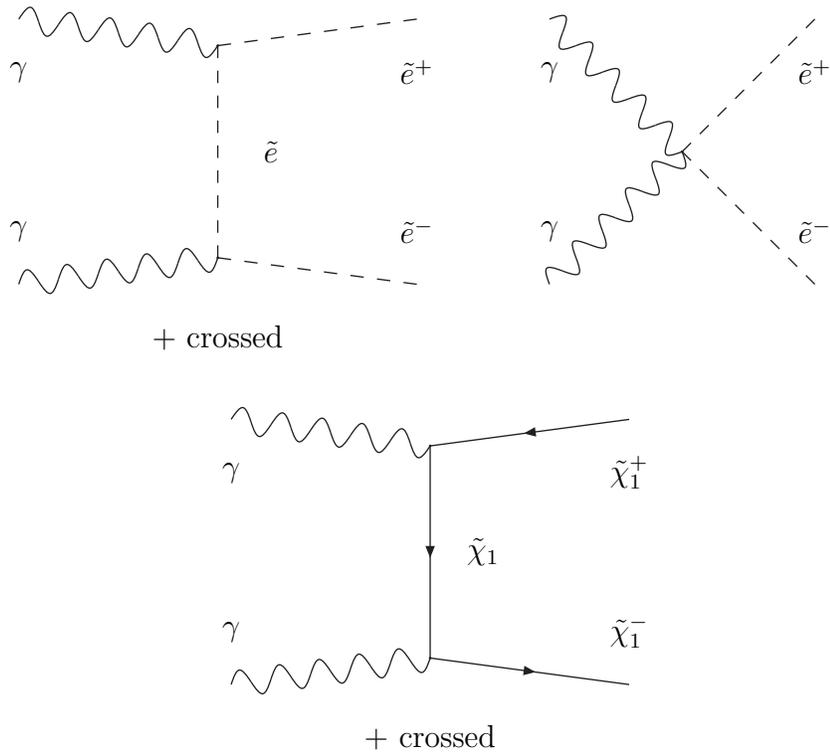

\begin{figure}[t]
\centerline{
\begin{picture}(554,504)(0,0)
\put(0,0){\strut\epsffile{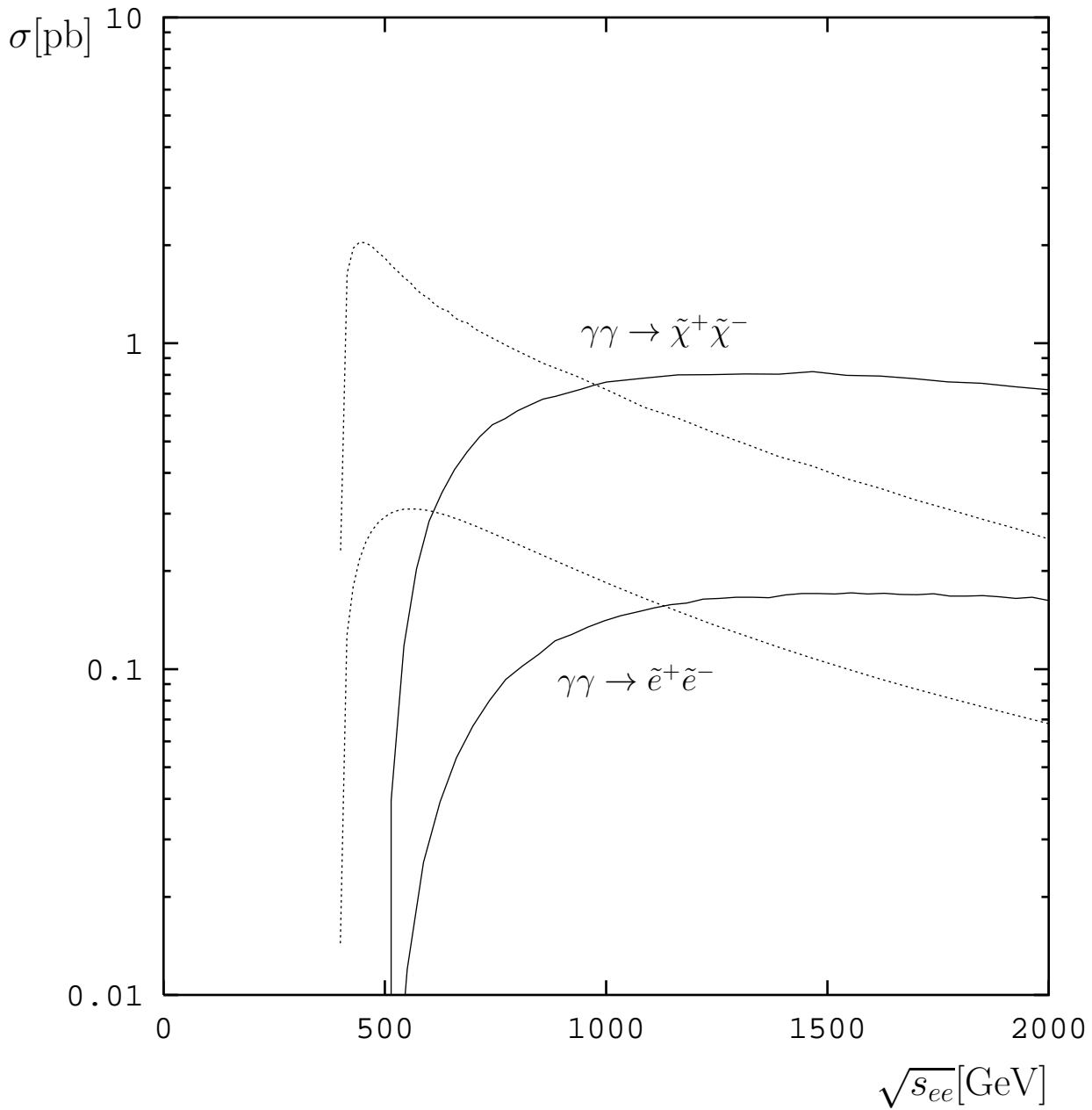}}
\put( 70.8,468.9){\makebox(0,0)[tr]{\Large$\sigma[\PB]$}}
\put(480.8, 19.1){\makebox(0,0)[tr]{\Large$\sqrt{s_{ee}}[\GEV]$}}
\put(280,330){\makebox(0,0)[bl]{{\large$\gamma\gamma\to\tilde\chi^+\tilde
\chi^-$}}}
\put(270,180){\makebox(0,0)[bl]{{\large$\gamma\gamma\to\tilde e^+\tilde e^-$}}}
\end{picture}
}
\caption[dummy]{Energy dependence of the pair-production \xs s
of 200 GeV \sel s and \co s.
The dotted curves are obtained
for monochromatic photons taking $\sqrt{s_{\gamma\gamma}}=\sqrt{s_{ee}}$.
The full curves result from the convolution
of the $\gamma\gamma$ \xs s with the Compton photon spectrum
 (\protect\ref{spectrum}).}
\label{f-1}
\end{figure}

To lowest order,
\sel s are pair-produced in photon-photon scattering
according to the Feynman diagrams of
Fig.~\ref{f1} (top).
The production \xs\ only depends on the mass and charge of the \sel.
We show, in Fig.~\ref{f-1}, the dependence of the integrated \xs\
on the collider energy $\sqrt{s_{ee}}$ for a 200 GeV \sel.
Results are given for monochromatic photon beams
with the nominal energy $\sqrt{s_{\gamma\gamma}}=\sqrt{s_{ee}}$,
and for photon beams with the energy spectrum (\ref{spectrum}).
In the latter case,
the \xs\ is given by the convolution formula
\begin{equation}
\sigma(s_{ee})=\int_0^{y_{max}} \!\!dy_1\int_0^{y_{max}}
        \!\!dy_2\ P(y_1)P(y_2)\sigma(s_{\gamma\gamma})
        \theta\left(y_1y_2s_{ee}-4m_{\tilde e}^2\right)~.
\label{fold}
\end{equation}
The remarkable change in the energy dependence of the \xs s
induced by the convolution over the Compton photon energy spectrum
is easily understood.
Close to the kinematic threshold of pair production,
only the most energetic of the Compton photons can contribute.
Hence the \xs\ is substantially reduced.
At higher energy,
more Compton photons contribute
and those photon pairs
whose \cm\ energy is just above threshold
have the highest \xs\ for pair-producing \sel s.
Therefore,
the folded \xs\ exceeds the \xs\ expected with monochromatic beams.

Since the \sel s decay by weak interactions
their width is typically much smaller than their mass
($\Gamma_{\Sel}/m_{\tilde e}\sim1\div.1\%$).
It is therefore safe to use the \nwa\ in the calculations.
We focus on the decay channel
\begin{equation}
        \Sel\to e^\pm\No
\label{seldecay}
\end{equation}
where the lightest \no\ \No\ is assumed to be the \lsp\
and therefore stable.
The \br\ for this decay is a complicated function
of all the masses and mixings in the gaugino-higgsino sector
\cite{LC1,LC2,LC5}
which themselves are very dependent
on the choice of the \susy\ parameters.
Later,
we shall explore systematically the accessible region of the parameter space.
However,
for definiteness
and in order to determine an optimized set of kinematical cuts,
we choose as a rather favourable scenario
\begin{eqnarray}
        \tan\beta &=& 4 \nonumber \\
        \mu &=& -400 \GeV \label{scenario} \\
        M_2 &=& 300 \GeV \nonumber
\end{eqnarray}
where $\tan\beta=v_2/v_1$ is the ratio of the Higgs \vev s,
and $\mu$ and $M_2$ are the soft \susy\ breaking mass parameters
associated with the higgsinos and the $SU(2)_L$ gauginos, respectively.
The $U(1)_Y$ gaugino mass parameter $M_1$
is assumed to evolve from the common value $M_1=M_2$ at the GUT scale
according to the relevant \rg\ equation
so that
$M_1=5/3\,M_2\tan^2\theta_w$,
where $\theta_w$ is the weak mixing angle.
For simplicity,
we also assume all sleptons to have the same mass,
and to be much lighter than the strongly interacting squarks and
gluinos\footnote{This last
condition will only become relevant in the next section
when we deal with the decay of \co s.}
\begin{equation}
        m_{\tilde\ell_L}=m_{\tilde\ell_R}=m_{\Snu_{\tilde\ell}}
        \ll m_{\tilde q},m_{\tilde g}~.
\label{assumption}
\end{equation}
In particular,
if $m_{\tilde e}=m_{\tilde\mu}$
all what is said about \sel\ production and decay
is also true for the production of smuons
and their subsequent decay into muons and invisible particles.

The \br\ for a \sel\ to decay into an electron and the lightest \no\
is 100\%\ if the \sel\ is lighter than all other \no s and \co s.
If the \sel\ is also kinematically allowed to decay into other channels,
this \br\ depends on
the masses of the involved particles
and the gaugino content of the \no s and \co s.
Typically,
the left-\sel\
({\em i.e.} the partner of the left-handed electron)
has a lower \br\ for the decay (\ref{seldecay})
than the right-\sel,
because the latter cannot decay into \co s.
A more detailed discussion of the decay patterns
can be found in Refs \cite{LC14,LC21}.

For the choice of parameters (\ref{scenario})
and a \sel\ mass of 300 GeV,
the left- and right-\sel s decay, respectively,
95\%\ and 100\%\ of the time
into the channel (\ref{seldecay}).
The left-\sel\ also decays with a 5\%\ \br\ into
$\tilde e^\pm_L\to\nu_e\Co$.
If $M_2$ is lowered to 150 GeV,
keeping all other parameters fixed,
one instead predicts
$BR(\tilde e_L\to e_L\No)=16\%$
while the \br\ of the right-\sel\ remains unaffected.
The integrated \xs\ of the $e^+e^-+\mpT$ signal
from \sel\ pair-production
is shown as a function of the collider energy in Fig.~\ref{f2}
for three different \sel\ masses.

Note that because of assumption (\ref{assumption})
and (s)lepton universality in the \mssm,
smuons and staus are pair-produced at the same rate as \sel s.
The $\tilde\tau^+\tilde\tau^-$ pairs
can also give rise to an $e^+e^-+\mpT$ signal
with the \br\
$
BR(\tilde\tau^+\tilde\tau^-\to e^+e^-+\mpT) =
BR(\tilde e^+\tilde e^-\to e^+e^-+\mpT)
BR(\tau\to e\nu_e\nu_\tau)^2
$.
There is thus an additional 3\%\ contribution to the
integrated
\susic\ signal
from stau production.
However,
since this a small correction
and since the \trm\ of these \ep\ pairs is somewhat degraded,
we do not consider it further.

%  #] selectrons:
%  #[ charginos:

\section{Production and Decay of Charginos}

To lowest order,
\co s are pair-produced in photon-photon scattering
as depicted in Fig.~\ref{f1} (bottom).
The dependence on the collider energy of the integrated \xs\
for producing 200 GeV \co s
is shown in Fig.~\ref{f-1}
for monochromatic and Compton back-scattered photons.
Again,
the convolution over the Compton photon energy spectrum
significantly modifies the energy dependence of the \xs.

As in the case of \sel s,
it is sufficient to use the \nwa\
in estimating the \xs\ for
$\gamma\gamma\to\tilde\chi_1^+\tilde\chi_1^-\to e^+e^-+\mpT$.
We focus on three different decay channels
which all ultimately yield the same final state:
{\arraycolsep0cm
\renewcommand{\arraystretch}{0}
\begin{equation}
        \Co \to
        \begin{array}[t]{ll}
                e^\pm&\Snu_e\\
                     &\hra\nu_e\No
        \end{array}
\label{codecay1}
\end{equation}
\begin{equation}
        \Co \to
        \begin{array}[t]{ll}
                \Sel^\pm\nu_e\\
                \hra e^\pm\No
        \end{array}
\label{codecay2}
\end{equation}
\begin{equation}
        \Co \to
        \begin{array}[t]{ll}
                W^\pm\No\\
                \hra e^\pm\nu_e
        \end{array}
\label{codecay3}
\end{equation}}
In principle three-body decays \cite{LC22}
should also be included in this list.
However,
their contribution is only sizable below
the two-body decay threshold
$m_{\tilde\chi_1^\pm}-m_{\tilde\chi_1^0}<m_W$,
and their neglect is of no consequence here.

As for $\Sel\to e\No$
the \br s of the decays (\ref{codecay1},\ref{codecay2},\ref{codecay3})
depend sensitively on the choice of the \susy\ parameters.
The most important of these parameters is the slepton mass.
Indeed,
if the lightest \co\ is lighter than the \sel\ or the \snu,
it can only decay into $W$'s via the reaction (\ref{codecay3})
and into charged Higgs bosons
via $\Co\to H^\pm\No$.
While $H^\pm\to e\nu_e$ is strongly suppressed,
the $W$ decays into an electron and neutrino,
however with a \br\ of only about 10\%.
Since this decay has to occur twice,
the total \br\ into an \ep\ pair is 1\%,
and the signal is hopelessly weak
(in principle, though, it may be increased by a factor 4
by also looking for $e\mu$ and $\mu\mu$ signals).
On the other hand,
if the \co\ is kinematically allowed to decay into sleptons,
it will preferably do so.
In this case,
because of (s)lepton universality
and because we assumed in (\ref{assumption})
that all sleptons have the same mass,
a \co\ will decay with equal probability into the three lepton families.
The \br s of the reactions (\ref{codecay1}) and (\ref{codecay2})
turn out to be almost equal
and roughly 17\%\ each
in the relevant region of the \susy\ parameter space,
while the decay into the $W$ channel is inhibited.
Nevertheless,
since the \co\ is then heavy,
its production \xs\ is low and
the expected $e^+e^-+\mpT$ signal remains weak.
This can be seen in Fig.~\ref{f2},
where we have plotted the energy dependence of the signal \xs\
for the \co\ channel.
In accordance with the scenario (\ref{scenario}),
the \co\ mass is approximately 290 GeV,
and we have considered a \sel\ mass of 200 GeV.
The result is approximately the same for
$m_{\tilde\chi_1^0}<m_{\tilde e}\lsim250$ GeV.
If the sleptons are heavier than 290 GeV
the expected signal \xs\ is approximately ten times lower.

Note that when a \co\ decays into a tau lepton,
the latter can also subsequently decay into electrons.
This provides an additional 38\%\ contribution to the \susic\ $e^+e^-+\mpT$
 signal,
which we take into account.
In contrast,
the heavier \co\ {\mbox{$\tilde\chi^\pm_2$}}
does not contribute significantly to the \susy\ signal.
Indeed,
for a sizable region of the \susic\ parameter space
its mass is much higher than the mass of the lighter \co\ \Co\
and its branching ratio into a lepton and invisible particles
is much less than one.

If one sticks to the assumption (\ref{assumption})
that the \sel\ and \snu\ have similar masses,
away from threshold the $e^+e^-+\mpT$ signal originating from the \co\
 production
is about an order of magnitude weaker
than the signal from \sel\ production.
This remains true for a large portion of the \susy\ parameter space.
It is only in very special circumstances,
for example, when $m_{\Snu} < m_{\tilde\chi_1^\pm} < m_{\tilde e}$,
that the \co\ signal dominates the \sel\ signal.
We do not consider this unlikely possibility here
and in the following we will optimize the cuts for the \sel\ signal only.

%  #] charginos:
%  #[ backgrounds:

\section{Backgrounds from the Standard Model}

The main \sm\ background processes
which also lead to an \ep\ pair
(and unobserved particles)
are the following:
{\arraycolsep0cm
\renewcommand{\arraystretch}{0}
\begin{equation}
        \gamma\gamma \to e^+e^-(\gamma)
\label{ep}
\end{equation}
\begin{equation}
        \gamma\gamma \to
        \begin{array}[t]{ll}
                e^+e^-&Z^0\\
                      &\hra\nu\bar\nu
        \end{array}
\label{eez}
\end{equation}
\begin{equation}
        \gamma\gamma \to
        \begin{array}[t]{ll}
                e^\pm\nu&W^\mp\\
                      &\hra e^\mp\bar\nu
        \end{array}
\label{enw}
\end{equation}
\begin{equation}
        \gamma\gamma \to
        \begin{array}[t]{ll}
                \tau^+&\tau^-\\
                      &\hra e^-\bar\nu_e\nu_\tau\\
                \makebox[0cm][l]{$\hra e^+\nu_e\bar\nu_\tau$}
        \end{array}
\label{tt}
\end{equation}
\begin{equation}
        \gamma\gamma \to
        \begin{array}[t]{ll}
                W^+&W^-\\
                      &\hra e^-\bar\nu_e\\
                \makebox[0cm][l]{$\hra e^+\nu_e$}
        \end{array}
\label{ww}
\end{equation}
\begin{equation}
        \gamma\gamma \to
        \begin{array}[t]{ll}
                W^+&W^-\\
                   &\hra e^-\bar\nu_e\\
                \makebox[0cm][l]{$\begin{array}[t]{ll}
                                        \hra&\tau^+\nu_\tau\\
                                                       &\hra
e^+\nu_e\bar\nu_\tau
                                  \end{array}$}
        \end{array}
\label{wt}
\end{equation}}

The most frequent process by far is the u- and t-channel electron-positron
pair-pro\-duc\-tion (\ref{ep}),
with a total \xs\ of the order of several hundred picobarn.
Nevertheless,
this background is also the easiest to eliminate.
Indeed,
here the \ep\ pairs are produced in a plane which contains the beam axis.
Therefore,
this background is totally eliminated
if we only consider acoplanar \ep\ events where
\begin{equation}
        ||\phi(e^+)-\phi(e^-)|-180^\circ| > 2^\circ~.
\label{phicut}
\end{equation}
Here, $\phi$ is the azimuthal angle
with respect to the beam axis.
By the requirement (\ref{phicut})
we exclude all \ep\ pairs which lie
on opposite sides within a wedge of $2^\circ$
whose axis is the beam axis.
Since the current detector angular resolution exceeds 3 mrad,
this is a conservative cut
which should also take care of all the unresolved Bremsstrahlung photons.

\begin{figure}[t]
\setepsfscale{.4}
\unitlength .4bp
\centerline{
\begin{picture}(554,554)(0,0)
\put(0,0){\strut\epsffile{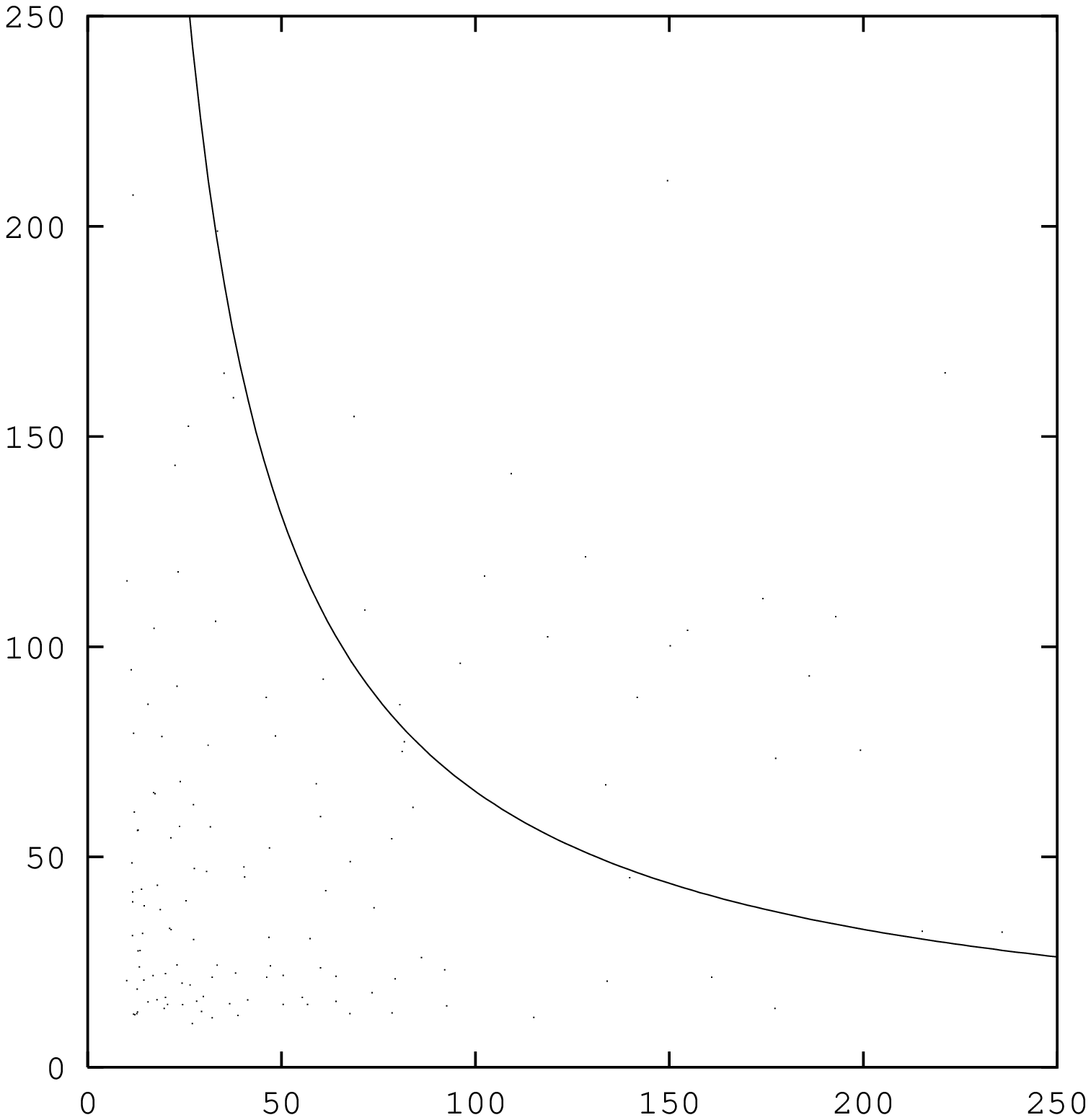}}
\put( 60.8,468.9){\makebox(0,0)[tr]{\shortstack[r]{$\pT(e^-)$\\$[\GEV]$}}}
\put(480.8, 19.1){\makebox(0,0)[tr]{$\pT(e^+)[\GEV]$}}
\put(450,450){\makebox(0,0)[tr]{{\Large$Z$}}}
\end{picture}
\begin{picture}(554,554)(0,0)
\put(0,0){\strut\epsffile{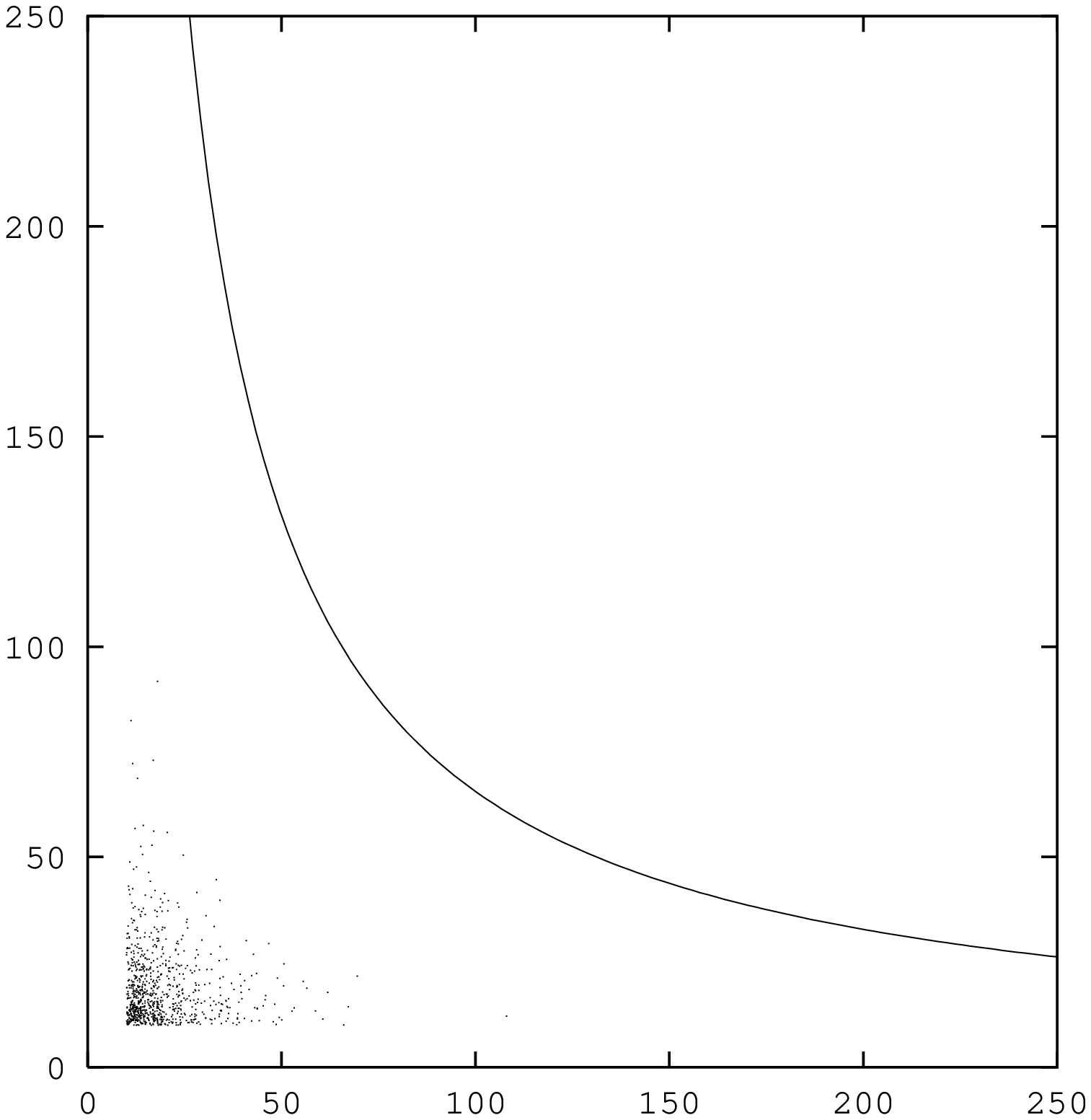}}
\put( 60.8,468.9){\makebox(0,0)[tr]{\shortstack[r]{$\pT(e^-)$\\$[\GEV]$}}}
\put(480.8, 19.1){\makebox(0,0)[tr]{$\pT(e^+)[\GEV]$}}
\put(450,450){\makebox(0,0)[tr]{{\Large$\tau$}}}
\end{picture}
}
\centerline{
\begin{picture}(554,554)(0,0)
\put(0,0){\strut\epsffile{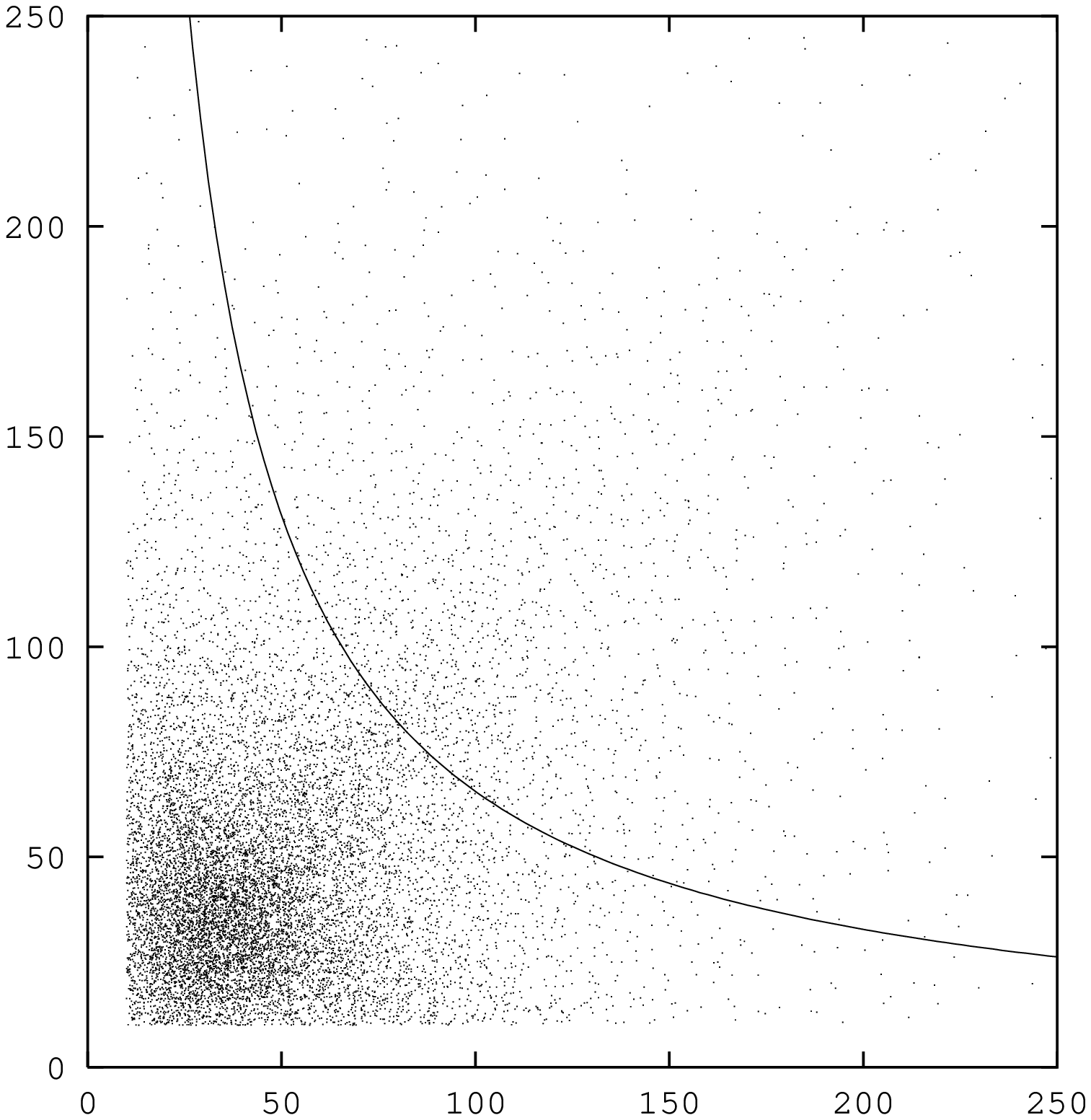}}
\put( 60.8,468.9){\makebox(0,0)[tr]{\shortstack[r]{$\pT(e^-)$\\$[\GEV]$}}}
\put(480.8, 19.1){\makebox(0,0)[tr]{$\pT(e^+)[\GEV]$}}
\put(450,450){\makebox(0,0)[tr]{{\Large$W$}}}
\end{picture}
\begin{picture}(554,554)(0,0)
\put(0,0){\strut\epsffile{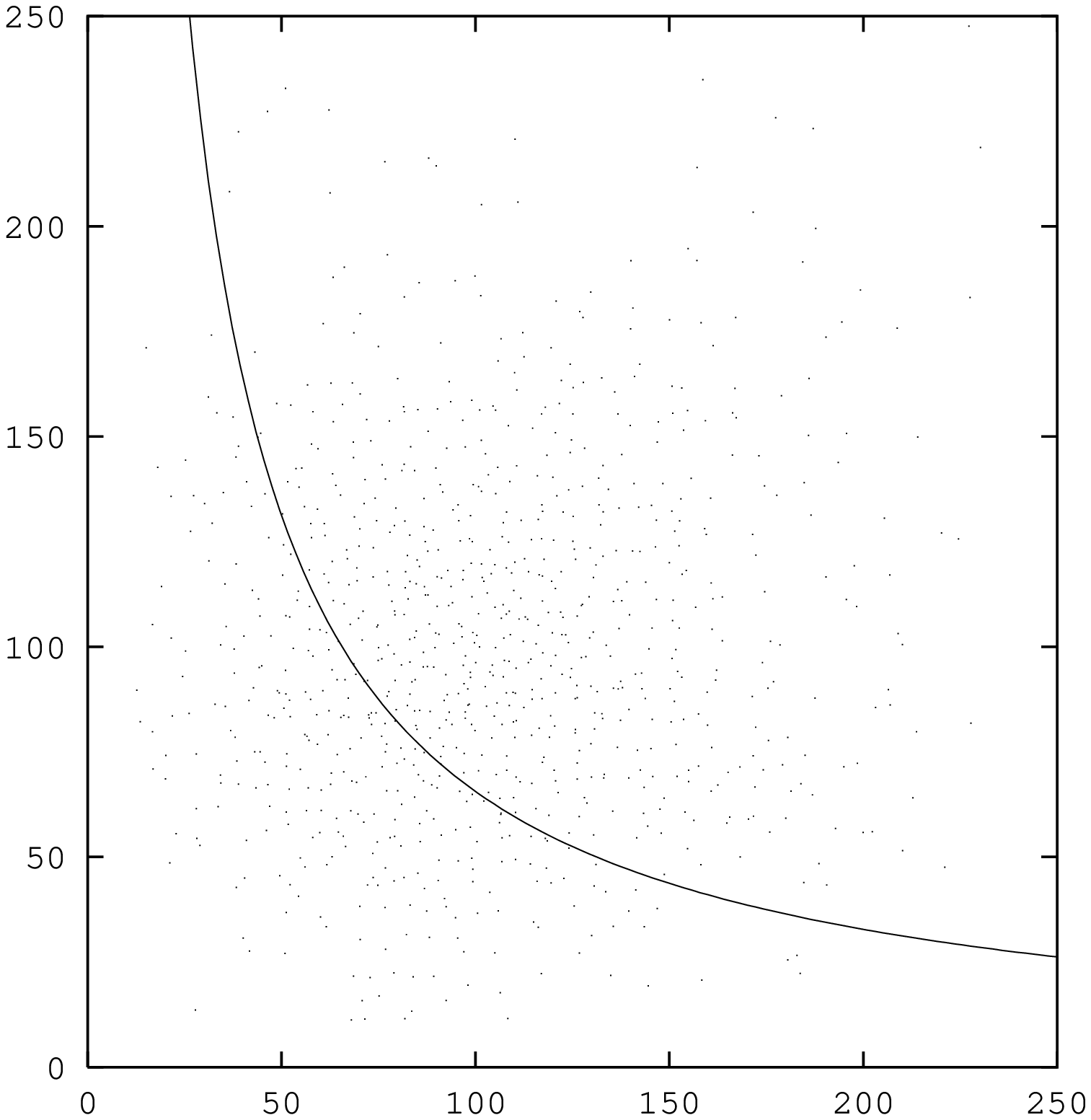}}
\put( 60.8,468.9){\makebox(0,0)[tr]{\shortstack[r]{$\pT(e^-)$\\$[\GEV]$}}}
\put(480.8, 19.1){\makebox(0,0)[tr]{$\pT(e^+)[\GEV]$}}
\put(450,450){\makebox(0,0)[tr]{{\Large$\tilde{e}$}}}
\end{picture}
}

\setepsfscale{1}
\unitlength 1bp

\caption[dummy]{Scatter plot of the electron and positron \trm\
at a 1 TeV collider
in the $Z$, $\tau$, $W$ and \Sel\ channels
(respectively Eqs (\ref{eez}), (\ref{tt}), (\ref{ww}) and (\ref{seldecay})).
The \sel\ mass has been set to $m_{\tilde e} = 300 \GeV$.
The remaining \susy\ parameters have been chosen according to the scenario
 (\ref{scenario}).
The set of cuts (A)
has been implemented,
%Every dot is equivalent to 50 ab.
and the region below the hyperbola is
excluded when the cut (\protect\ref{hyp}) is used.}
\label{f3}
\end{figure}

However,
this cut fails to eliminate events which arise
when the electron emits a $Z^0$,
which subsequently decays into neutrinos (\ref{eez}).
Nevertheless,
the cross section is only sizeable when
either the electron or the positron or both
have small \trm\ \pT.
Hence,
a moderate cut,
such as
\begin{equation}
        \min(\pT(e^+),\pT(e^-)) > 10 \GeV,
        \label{ptcut}
\end{equation}
sufficiently suppresses also this source of background.
The selection criteria (\ref{phicut},\ref{ptcut})
have been implemented in Figs \ref{f3}
where we show how the \trm\ of the electrons and positrons
are distributed at a 1 TeV collider.
Every event on this scatter plot carries a weight of 50 attobarn.

The $e\nu_eW$ background (\ref{enw})
is more difficult to compute
because it includes the off-shell part of the $W^+W^-$ reaction (\ref{ww}).
A careful study \cite{LC20}, though,
reveals that the interference between the resonant and radiative diagrams
is small.
The latter is expected to yield a contribution of the same order
as the $e^+e^-Z^0$ background (\ref{eez}).

\begin{figure}[t]
\centerline{
\begin{picture}(554,504)(0,0)
\put(0,0){\strut\epsffile{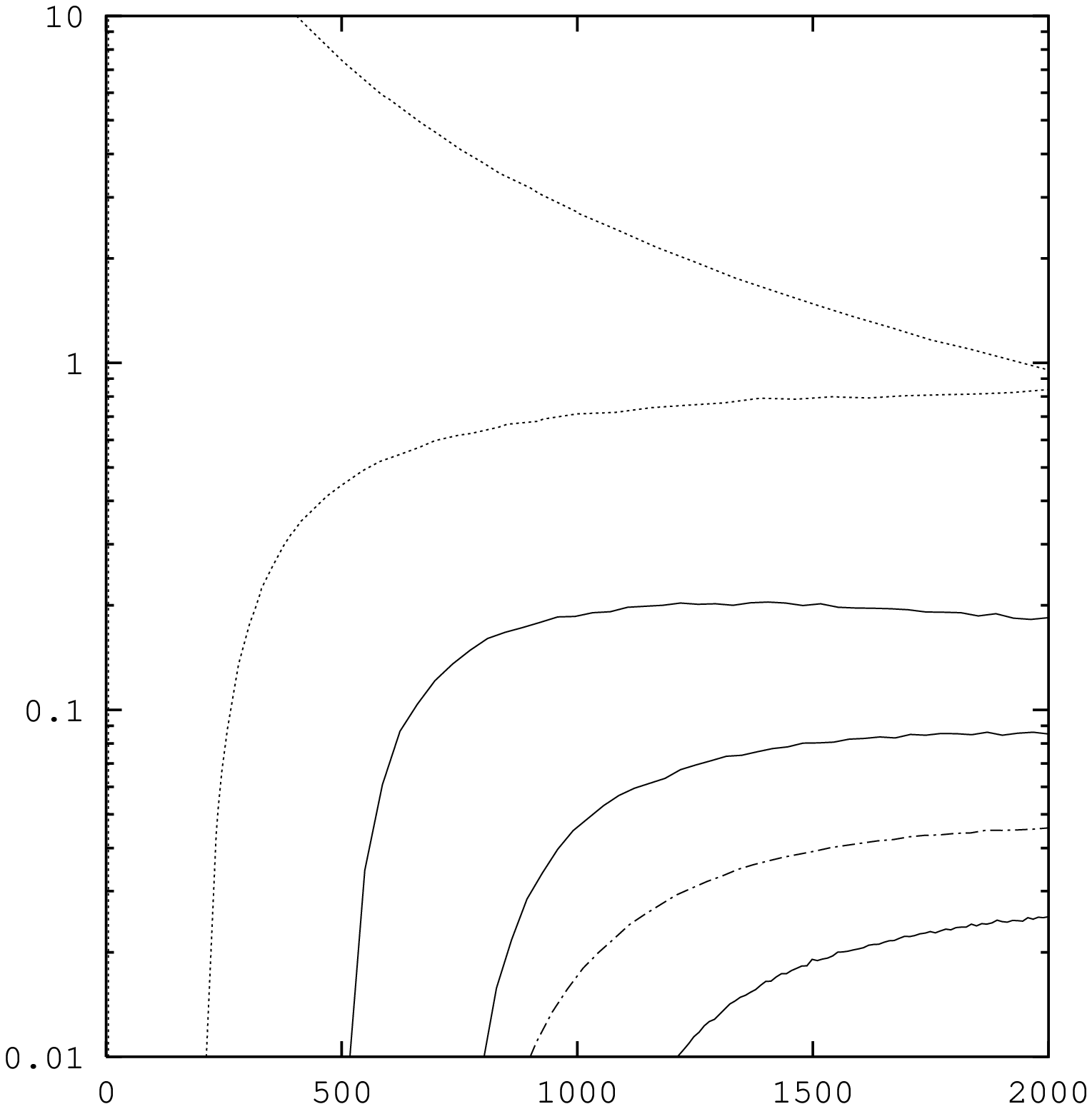}}
\put( 70.8,468.9){\makebox(0,0)[tr]{\Large$\sigma[\PB]$}}
\put(480.8, 19.1){\makebox(0,0)[tr]{\Large$\sqrt{s_{ee}}[\GEV]$}}
\put(272,400){\makebox(0,0)[bl]{{\Large$\tau^\pm$}}}
\put(260,303){\makebox(0,0)[br]{{\Large$W^\pm$}}}
\put(292,224){\makebox(0,0)[tl]{{\Large$\tilde{e}^\pm$}$(200\GeV)$}}
\put(339,166){\makebox(0,0)[tl]{{\Large$\tilde{e}^\pm$}$(300\GeV)$}}
\put(386,086){\makebox(0,0)[tl]{{\Large$\tilde{e}^\pm$}$(400\GeV)$}}
\put(362,126){\makebox(0,0)[tl]{{\Large${\tilde{\chi}}_1^\pm$}$(290\GeV)$}}
\end{picture}
}
\caption[dummy]{Energy dependence of the signal \xs s for
$\gamma\gamma\to\tilde e^+\tilde e^-\to e^+e^-+\mpT$
and
$\gamma\gamma\to\tilde\chi^+\tilde\chi^-\to e^+e^-+\mpT$
assuming scenario (\protect\ref{scenario}).
The \co\ prediction is obtained
for sleptons lighter than 250 GeV.
The background \xs s for
$\gamma\gamma\to W^+W^-\to e^+e^-+\mpT$
and
$\gamma\gamma\to\tau^+\tau^-\to e^+e^-+\mpT$
are also shown.}
\label{f2}
\end{figure}

The \trm\ cut (\ref{ptcut}) also almost entirely
eliminates the \ep\ pairs
originating from $\tau$ production (\ref{tt}).
In Fig.~\ref{f2}\ we display the \xs\ for the channel
$\gamma\gamma\to\tau^+\tau^-\to e^+e^-+$neutrinos.
The scatter plot in Fig.~\ref{f3}\
reveals that the decay electrons and positrons
clutter in the very low \trm\ region.
This is a manifestation of the u- and t-channel poles
in the $\tau$ production amplitude,
leading to a distribution of the $\tau$-leptons
(and hence the decay electrons and positrons)
which strongly peaks in the forward and backward direction.
%D'amour, vos beaux yeux, belle Marquise, me font mourrir!

The most important background consists of \ep\ pairs
which arise from the decay of
on-shell
$W^+W^-$ pairs (\ref{ww}) \cite{LC7}.
In Fig.~\ref{f2}\ we display the \xs\ for the channel
$\gamma\gamma\to W^+W^-\to e^+e^-+$neutrinos.
As is shown in Fig.~\ref{f3},
some of these \ep\ pairs can be produced at very high \trm,
but most of them populate the region close to
$\pT(e^+)\approx\pT(e^-)\approx40\GeV$.

A non-negligible addition to this background
consists of electrons or positrons
originating from the decay of a $\tau$
which itself is a decay product of one of the $W$'s (\ref{wt}).
When no cuts are applied,
it increases the background from $W^+W^-$ production by 38\%.
However,
when the \trm\ cut (\ref{ptcut}) is applied,
this contribution drops to less than 20\%\
because the \trm\ of the $e^\pm$ undergo a further degradation
in the additional cascade.

Of course,
there are more sources of background,
but all are significantly smaller than the ones already considered.

%  #] backgrounds:
%  #[ results:

\section{Results}
The search for \susy\ signals
of the kind considered here
is facilitated by the fact that
the \sm\ backgrounds
can be evaluated theoretically to a high accuracy.
The validity of these calculations can even be checked experimentally
for $W^+W^-$ and $\tau^+\tau^-$ pairs yielding $e\mu$ events,
which cannot be obtained from \sel\ pair production
(\co\ pair production, though, could also yield such $e\mu$ events,
but we have seen that this is generally a negligible contribution).
Any statistically relevant deviation from the \sm\ prediction
could thus be a sign for \susy.
The task is then to discriminate this hypothesis against other interpretations.

\begin{figure}[t]
\setepsfscale{.4}
\unitlength .4bp
\centerline{
\begin{picture}(554,554)(0,0)
\put(0,0){\strut\epsffile{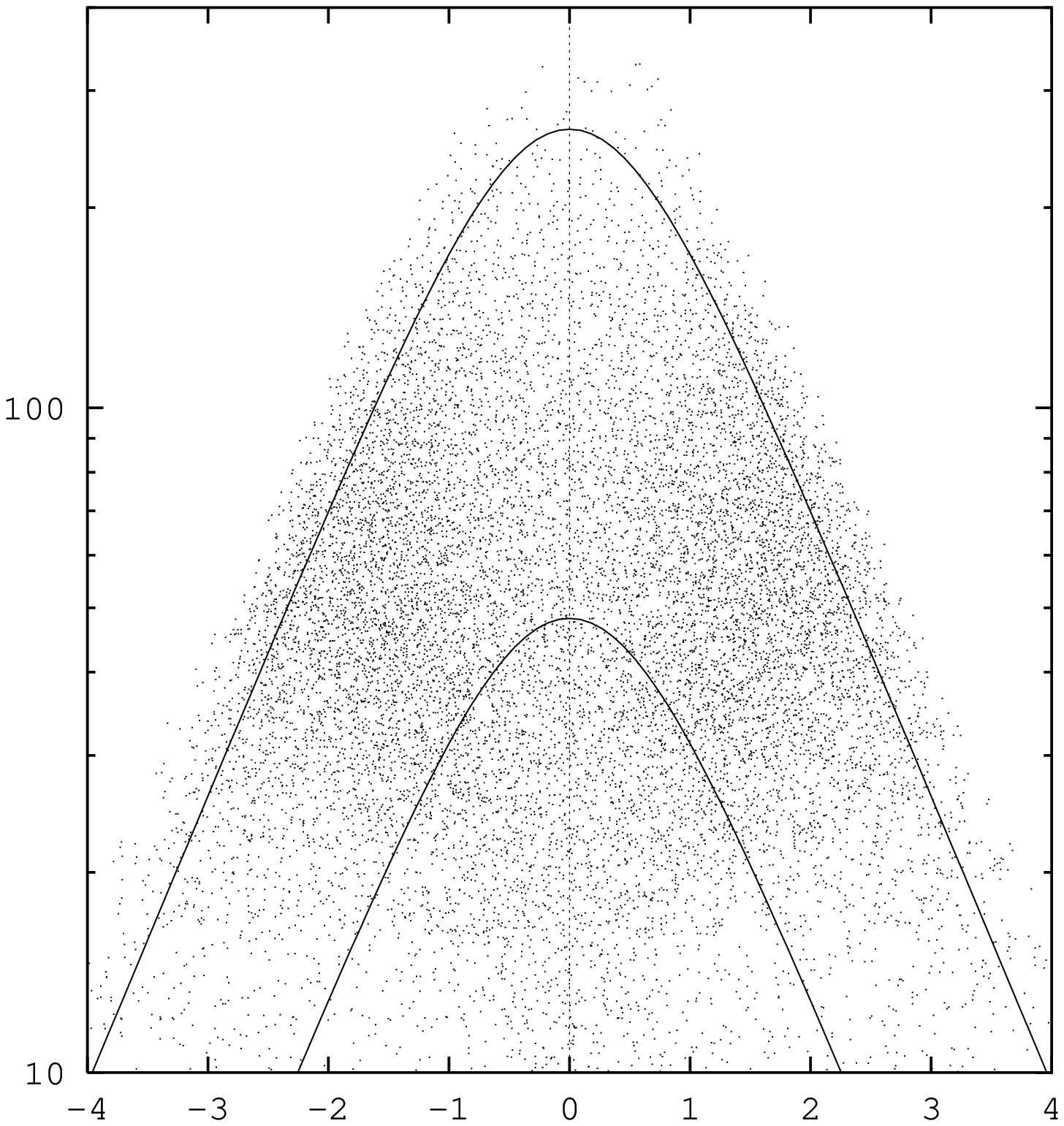}}
\put( 60.8,468.9){\makebox(0,0)[tr]{\shortstack[r]{$\pT(e^\pm)$\\$[\GEV]$}}}
\put(480.8, 19.1){\makebox(0,0)[tr]{$\eta(e^\pm)$}}
\put(450,450){\makebox(0,0)[tr]{{\Large$W$}}}
\end{picture}
\begin{picture}(554,554)(0,0)
\put(0,0){\strut\epsffile{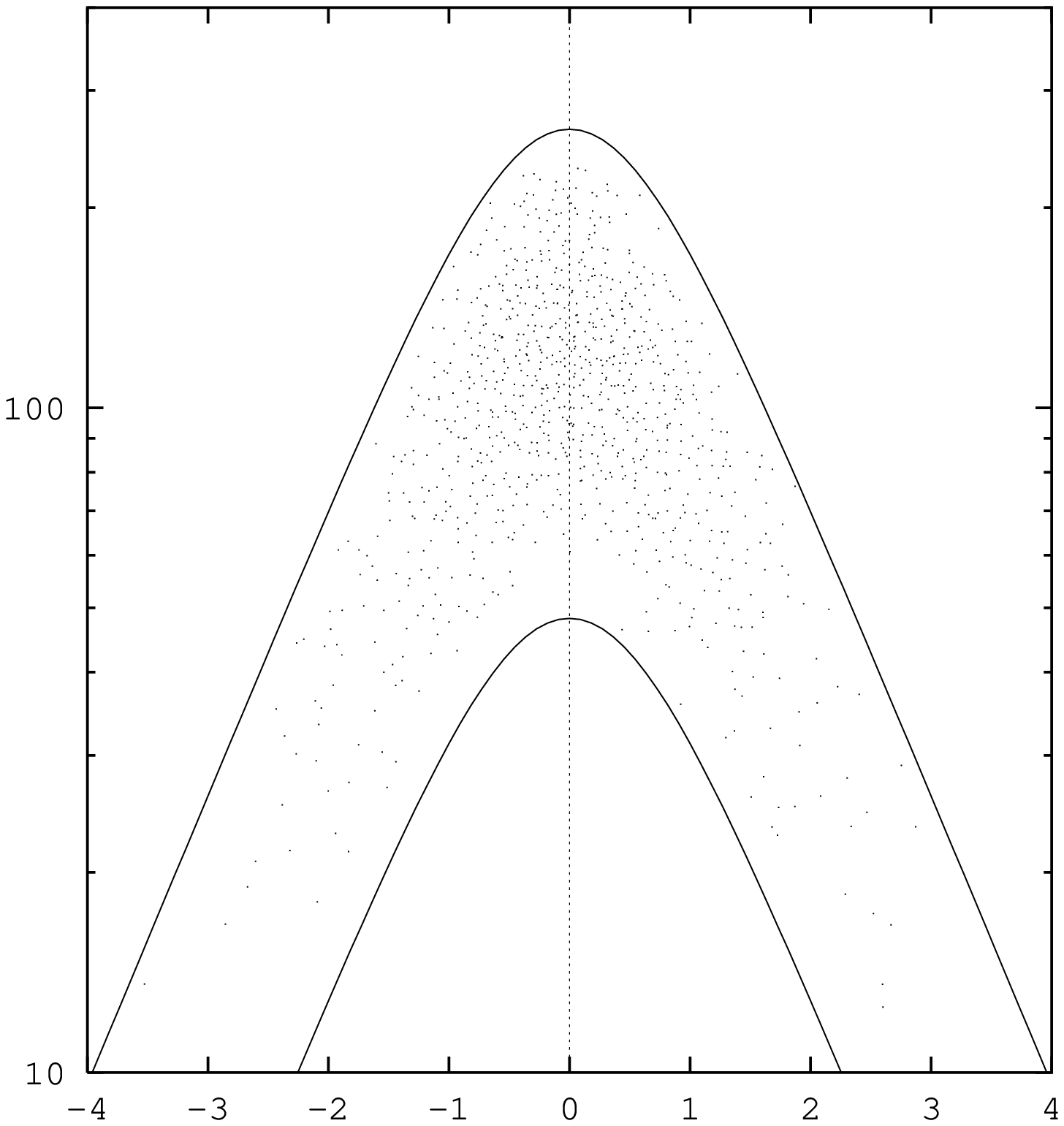}}
\put( 60.8,468.9){\makebox(0,0)[tr]{\shortstack[r]{$\pT(e^\pm)$\\$[\GEV]$}}}
\put(480.8, 19.1){\makebox(0,0)[tr]{$\eta(e^\pm)$}}
\put(450,450){\makebox(0,0)[tr]{{\Large$\tilde{e}$}}}
\end{picture}
}

\setepsfscale{1}
\unitlength 1bp

\caption[dummy]{Distribution of the decay leptons' \trm\ versus rapidity
at a 1 TeV collider
in the $W$ and \Sel\ channels
(respectively Eqs (\ref{ww}) and (\ref{seldecay})).
The \sel\ mass has been set to $m_{\tilde e} = 300 \GeV$.
The remaining \susy\ parameters have been chosen according to the scenario
 (\ref{scenario})
and the set of cuts (A)
has been implemented.}
%Every dot is equivalent to 50 ab.}
\label{fboom}
\end{figure}

To obtain an appreciable signal to background ratio,
it is necessary to impose further cuts
to reduce the $W$ background (\ref{ww}).
As can be seen from Figs \ref{f3},
for not too heavy \sel s,
this is the only dangerous background left
after the cuts (\ref{phicut},\ref{ptcut}) have been implemented.
We suggest the following low \trm\ and high rapidity cuts:
\begin{equation}
   \pT(e^+)\pT(e^-) > m_W^2~,
\label{hyp}
\end{equation}
\begin{equation}
\label{rap}
        |\eta(e^\pm)| < 1~.
\end{equation}
As is shown on the scatter plot of Fig.~\ref{f3}
for the choice of parameters (\ref{scenario}),
the \trm\ distribution of \ep\ pairs originating from 300 GeV \sel s
is peaked around
$\pT(e^+)\approx\pT(e^-)\approx100\GeV$
and the \susic\ signal is not exaggerately curtailed by the low \trm\ cut
 (\ref{hyp}).
Similarly,
few of the electrons or positrons are emitted at low angles.
This can be observed on the scatter plot of Fig.~\ref{fboom},
where we have displayed the correlation
between the rapidity and the \trm\ of the electron (or positron).
As in Fig.~\ref{f3},
events are distributed here with a weight of 50 attobarn.

If \susy\ has been discovered and the masses of the \sel\ and lightest \no\
are known,
or if the data analysis is performed with varying cuts,
one can even further enhance the signal to background ratio.
Indeed,
the energy of the signal electrons or positrons
is kinematically bounded from above and from below:
\begin{equation}
        E_e \in {E\over4} \left[1-{m^2_{\No}\over m^2_{\Sel}}\right]
                \left[1\pm\sqrt{1-{4m^2_{\Sel}\over E^2}}\right]
\label{eny}
\end{equation}
where $E=\sqrt{s_{ee}}x/(x+1)\approx.83\sqrt{s_{ee}}$
is the maximum attainable \cm\ energy
in the photon-photon collision.
These two boundaries are depicted by the boomerang curves in Fig.~\ref{fboom}.
Clearly,
a lot of $W$ background can be further eliminated
by imposing the energy cuts (\ref{eny}) on the electrons and positrons,
without affecting the signal.

\begin{figure}[t]
\centerline{
\begin{picture}(554,504)(0,0)
\put(0,0){\strut\epsffile{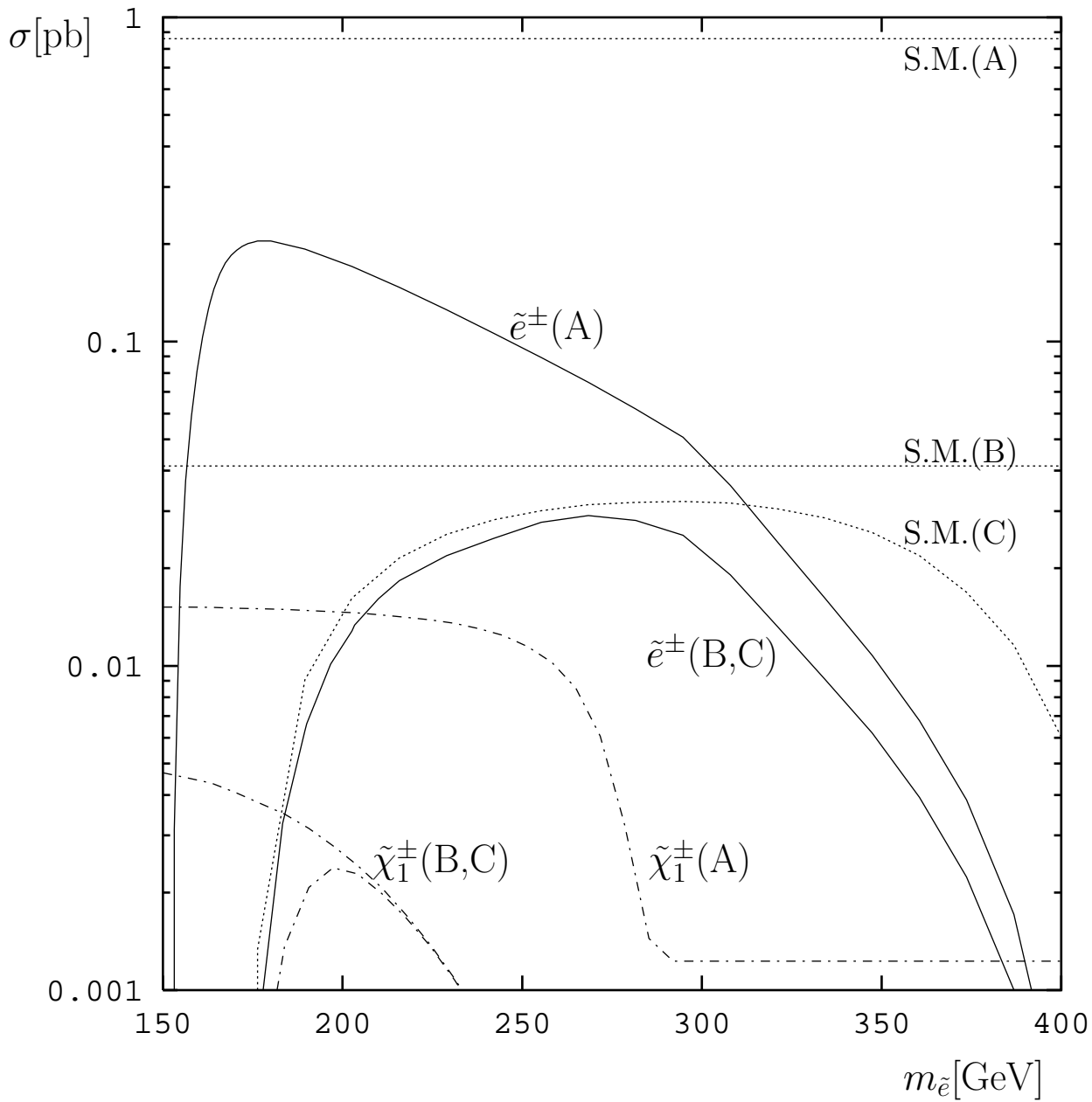}}
\put( 70.8,468.9){\makebox(0,0)[tr]{\Large$\sigma[\PB]$}}
\put(480.8, 19.1){\makebox(0,0)[tr]{\Large$m_{\tilde{e}}[\GEV]$}}
\put(420,457){\makebox(0,0)[tl]{{\large S.M.(A)}}}
\put(420,277){\makebox(0,0)[bl]{{\large S.M.(B)}}}
\put(420,242){\makebox(0,0)[bl]{{\large S.M.(C)}}}
\put(250,330){\makebox(0,0)[bl]{{\Large$\tilde{e}^\pm$(A)}}}
\put(366,202){\makebox(0,0)[tr]{{\Large$\tilde{e}^\pm$(B,C)}}}
\put(310,100){\makebox(0,0)[bl]{{\Large${\tilde{\chi}}_1^\pm$(A)}}}
\put(191,100){\makebox(0,0)[bl]{{\Large${\tilde{\chi}}_1^\pm$(B,C)}}}
\end{picture}
}
\caption[dummy]{Dependence on the \sel\ mass
of the \xs\ of the $e^+e^-+\mpT$ signals
at a 1 TeV collider.
The \sel\ and \co\ channels as well as the \sm\ background
are shown for the three sets of cuts (A,B,C) discussed in the text.
The \co\ mass is 290 GeV, in accordance with scenario
(\protect\ref{scenario}).}
\label{f4}
\end{figure}

In Fig.~\ref{f4}\ we show how the \sel\ and \co\ signals
compare to the \sm\ background
at a 1 TeV collider
as a function of the \sel\ mass.
The \susic\ signals and \sm\ backgrounds are displayed
for the different types of cuts discussed above:
\begin{itemize}
\item[(A)] only the acoplanarity and low \trm\ cuts (\ref{phicut},\ref{ptcut});
\item[(B)] the cuts (A) plus the \trm\ and rapidity cuts (\ref{hyp},\ref{rap});
\item[(C)] the cuts (B) plus the energy cuts (\ref{eny}).
\end{itemize}

We first comment on the process
$\gamma\gamma\to\tilde e^+\tilde e^-\to e^+e^-+\No\No$.
Since the maximum energy (\ref{eny}) of the electrons or positrons
originating from \sel s
is proportional to the difference of the masses
of the \sel\ and the \no\
the signal is lost with the \trm\ cut (\ref{ptcut})
if the \sel\ mass is close to 150 GeV
(the lightest \no\ mass,
according to the scenario (\ref{scenario})).
Because left-\sel s of more than 290 GeV
(the lightest \co\ mass,
according to the scenario (\ref{scenario}))
can also decay into \co s,
the signal drops noticeably beyond this \sel\ mass.
The \trm\ and rapidity cuts (\ref{hyp},\ref{rap})
reduce the \sm\ background by approximately a factor 20,
while the \sel\ signal is only slightly reduced
for $m_{\tilde e}\gsim250$ GeV.
At least for $m_{\tilde e}\lsim300$ GeV,
the mass dependent energy cut (\ref{eny})
leads to a further significant improvement of the signal to background ratio.

Turning to the process
$\gamma\gamma\to\tilde\chi^+\tilde\chi^-\to e^+e^-+\No\No\nu\bar\nu$
we note that the \xs\
including the set of cuts (\ref{phicut},\ref{ptcut})
is almost independent of the \sel\ mass,
except when the latter is close to the lightest \co\ mass,
that is 290 GeV.
Beyond this point the \co\ cannot decay anymore into
\sel-neutrino or electron-\snu\ pairs.
In that case,
the $e^+e^-+\mpT$ signal is only obtained
from the decay of \co s into \no s and $W$'s,
which subsequently decay with a \br\ of $BR(W\to e\nu_e)^2=1\%$
into electrons and neutrinos.
Because of this low branching ratio,
the \co\ channel yields very few \ep\ pairs for heavy \sel s.
If the \trm\ cut (\ref{hyp}) is applied,
the signal becomes even more suppressed
the closer the \co\ and slepton masses are.
The mass dependent energy cut (\ref{eny})
makes of course matters even worse.

\begin{figure}[t]
\centerline{
\begin{picture}(554,504)(0,0)
\put(0,0){\strut\epsffile{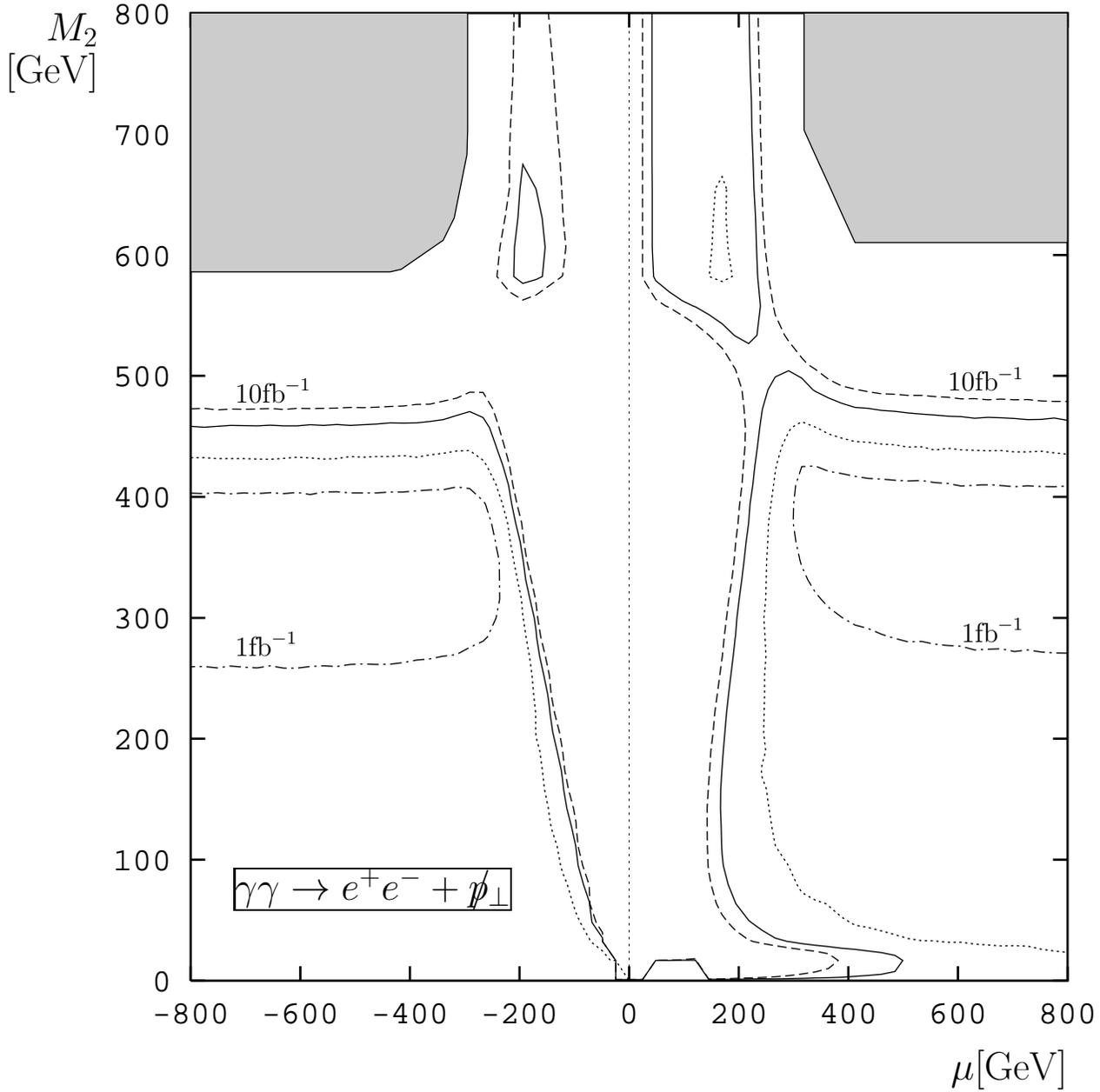}}
\put( 60.8,468.9){\makebox(0,0)[tr]{\Large\shortstack[r]{$M_2$\\$[\GEV]$}}}
\put(480.8, 19.1){\makebox(0,0)[tr]{\Large$\mu[\GEV]$}}
\put(120,190){\makebox(0,0)[bl]{$ 1 \FB^{-1}$}}
\put(120,300){\makebox(0,0)[bl]{$ 10 \FB^{-1}$}}
\put(462,196){\makebox(0,0)[br]{$ 1 \FB^{-1}$}}
\put(462,305){\makebox(0,0)[br]{$ 10 \FB^{-1}$}}
\put(120,80){\makebox(0,0)[bl]{\Large\frame{$\gamma\gamma\to e^+e^-+\mpT$}}}
\end{picture}
}
\caption[dummy]{Regions in the $(\mu,M_2)$ plane where the
\susic\ signal
due to a 300 GeV selectron
can be distinguished from the \sm\ background
at a $3\sigma$ confidence level.
The assumed collider energy is 1 TeV
and results are shown for 1, 2, 5 and 10 $\FB^{-1}$
of integrated luminosities.
The shaded areas,
in which $m_{\tilde e}<m_{\tilde\chi_1^0}$,
are excluded
and $\tan\beta=4$.}
\label{f5}
\end{figure}

In order to transcend the illustrative scenario (\ref{scenario})
and to show the dependence of the $e^+e^-+\mpT$ signal on the \susic\
 parameters,
we plot in Fig.~\ref{f5}
the limits of observability in the $(\mu,M_2)$ plane
for a 300 GeV \sel\
at a 1 TeV collider
for four integrated luminosities:
1, 2, 5 and 10 fb$^{-1}$.
For this,
we demand that the signal be
at least three events and
at least three standard deviations above the background's Poisson fluctuations:
\begin{equation}
        n_{SUSY} > 3\sqrt{n_{SM}}~.
\label{poisson}
\end{equation}
Note that these results are conservative
in the sense that
we only considered here the decay of \sel s into electrons/positrons
and the lightest \no.
Particularly for low values of $\mu$ or $M_2$,
cascade decays of the \sel\ are important
and are likely to contribute to the $e^+e^-+\mpT$ signal,
with a further degradation of the \trm\ though.

%  #] results:
%  #[ conclusions:

\section{Conclusions}

%This is the most excellent paper ever written about \susy\ and \lpc's.
Sleptons and \co s can be pair-produced
with sizeable rates
at linear electron-positron colliders operated in the photon-photon mode.
Although
the discovery potential in this mode cannot compete with the potential
of other reactions,
the prospects for complementary studies of \susy\ in $\gamma\gamma$ collisions
are particularly interesting.
This is because there is no model dependence
at the production level
so that the decay properties can be investigated in a clean way.
This may yield important information on the gaugino-higgsino sector.

A very promising signature for \sel\ pair production
consists of an acoplanar \ep\ pair.
The same signal can be obtained from pair-produced \co s,
but for most choices of \susic\ parameters
their \br s into electrons remain small.
As a consequence,
away from threshold,
the $e^+e^-+\mpT$ signal from \sel\ pair production
dominates by about one order of magnitude
the one expected from \co\ pair production.

The only significant \sm\ background
which remains after mild acoplanarity and \trm\ cuts
consists of \ep\ pairs resulting from the decay of pair-produced $W$'s.
This background can be reduced by a factor 20
with more stringent low \trm\ and high rapidity cuts.
A further energy cut,
which assumes the masses of the \sel\ and lightest \no\ to be known,
can even yield a signal to background ratio of one.
Of course,
all these results are trivially extended to the pair production of smuons
and their subsequent decay into $\mu^+\mu^-$ pairs and missing energy.

We conclude that high energy linear \ep\ colliders
operated in the $\gamma\gamma$ mode
provide novel possibilities
for \susy\ searches
with integrated luminosities as low as 1 fb$^{-1}$.
Particularly interesting results should be obtained
in conjunction with similar searches
in $e^\pm e^-$ \cite{LC1,LC11,LC18}
and $e^-\gamma$ \cite{LC10,LC16,LC19}
collisions.

\bigskip
\bigskip
\bigskip
F.C. wishes to express his gratitude to Michael Dine and Howie Haber
for their hospitality
at the U.C.S.C. Physics Department,
where part of this work was performed.
We also acknowledge partial support by CED Science Project No.
SCI-CT 91-0729.

%  #] conclusions:
%  #[ bibliograpy:

\bibliographystyle{unsrt}

%  #] bibliograpy:
\newpage
%  #[ figures:

\section{Figures}

%  #] figures:
%\listoffigures

\end{document}